\documentclass[
 aip,
 pop,
 reprint,
 amsmath,amssymb
]{revtex4-1}
\usepackage[english]{babel}

\usepackage{graphicx}
\usepackage{dcolumn}
\usepackage{bm}

\usepackage[utf8]{inputenc}
\usepackage[T1]{fontenc}
\usepackage{mathptmx}
\usepackage{etoolbox}
\usepackage{makecell}

\usepackage{colortbl}
\usepackage{siunitx}
\usepackage{colortbl}
\usepackage{xcolor}
\usepackage{multirow}
\usepackage{relsize} 
\usepackage{xurl}
\usepackage[normalem]{ulem}

\makeatletter
\def\@email#1#2{%
 \endgroup
 \patchcmd{\titleblock@produce}
  {\frontmatter@RRAPformat}
  {\frontmatter@RRAPformat{\produce@RRAP{*#1\href{mailto:#2}{#2}}}\frontmatter@RRAPformat}
  {}{}
}%
\makeatother

\begin{document}

\preprint{AIP/123-QED}

\title[BO of Mo-99 Production by LWFA Using Coupled PIC-MC Simulations]{
Bayesian Optimization of Molybdenum-99 Production by Laser Wakefield Acceleration Using Coupled PIC and Monte Carlo Simulations}
\author{B.S. Nunes}
\affiliation{Nuclear and Energy Research Institute, IPEN–CNEN, São Paulo 05508-000, Brazil.}
\email{brunosnunes@usp.br}

\author{N.D. Vieira Jr.}
\affiliation{Nuclear and Energy Research Institute, IPEN–CNEN, São Paulo 05508-000, Brazil.}

\author{M.S. Alva-Sánchez}
\affiliation{Graduate Program in Information Technology and Healthcare Management, Federal University of Health Sciences of Porto Alegre (UFCSPA), Porto Alegre 90050-170, Brazil.}

\author{A. Bonatto}
\affiliation{Graduate Program in Information Technology and Healthcare Management, Federal University of Health Sciences of Porto Alegre (UFCSPA), Porto Alegre 90050-170, Brazil.}
 
 \author{R.E. Samad}
 \affiliation{Nuclear and Energy Research Institute, IPEN–CNEN, São Paulo 05508-000, Brazil.}
 
\date{\today}

\begin{abstract}
This work applies Bayesian optimization to a loop composed of PIC simulations of laser electron acceleration and Monte Carlo (MC) simulations of bremsstrahlung-induced nuclear reactions, to maximize the production of molybdenum-99, the precursor of the most used radiopharmaceutical in nuclear medicine, metastable technetium-99. PIC and MC simulations are computationally intensive, and besides reducing the time spent, the Bayesian optimization coupling both simulations 
resulted in an improvement of an order of magnitude in the $^\text{99}$Mo yield over a previous work, in which the output of an optimization loop based solely on PIC simulations was used \textit{a posteriori} to estimate $^{99}\mathrm{Mo}$ production through a MC simulation.

\end{abstract}

\maketitle

\section{\label{sec:intro}Introduction}

The global radioisotope supply chain for nuclear medicine applications is highly dependent on the neutron-induced fission of uranium-235 ($^{235}$U) in research reactors\cite{Ge2023}, the main production route for molybdenum-99 ($^\text{99}$Mo, half-life of 66~\si{\hour}) \cite{Xu2024}, which decays to technetium-99m ($^{99\text{m}}$Tc, half-life of 6~\si{\hour}). $^{99\text{m}}$Tc is employed in more than 40 million single-photon emission computed tomography \cite{Varga2023} procedures every year, being the most widely used radioisotope globally. However, the demand on using low-enriched uranium---coupled with the limited number of reactors in operation and their advanced age---has exposed the fragility of this supply chain, exemplified by the global $^\text{99}$Mo shortage in 2009 \cite{VanNoorden2013}, when the simultaneous shutdown of the two major production reactors triggered a worldwide crisis, resulting in the cancellation of countless diagnostic procedures. Besides, the production is centralized in a small number of nuclear reactors, requiring a complex global distribution logistic, which is constrained by the locations of medical facilities and by the regulations governing the transport of radioactive materials.  This situation is a strong motivation for developing alternative, decentralized, accelerator-based routes for medical radioisotope production that avoid the use of enriched uranium \cite{Nunes2022}.

A promising alternative for $^\text{99}$Mo production involves the use of bremsstrahlung photons generated by high-energy electron beams to trigger the photonuclear reaction $^\text{100}$Mo($\gamma$, $n$)$^\text{99}$Mo \cite{Trknyi2018} on natural molybdenum ($^\text{nat}$Mo) targets. Electron beams with energies of several tens of MeV are efficient to activate this reaction \cite{Nunes2026}, and can be produced either by conventional accelerators or by compact plasma-based sources such as laser wakefield accelerators (LWFA) \cite{Tajima1979, Esarey2009}. In the LWFA process, an intense laser pulse propagates through an underdense plasma, exciting longitudinal electric fields that can reach gradients on the order of a few \si{\tera\eV/\m} \cite{Lai2023}, thereby accelerating electrons to relativistic energies within a few micrometers. When these electrons collide with a high-$Z$ converter, such as tantalum, they emit bremsstrahlung photons capable of inducing the photonuclear reaction that produces $^\text{99}$Mo from the naturally-occurring, stable isotope $^\text{100}$Mo \cite{Martin2017}. The overall efficiency of this multi-stage process is strongly influenced by the characteristics of the electron beam produced in the plasma stage, as well as by the geometry and material configuration of the converter–target assembly.

The blowout (or bubble) LWFA regime is known for producing high-energy electron beams with low energy spread and small emittance \cite{Gonsalves2019}. However, it requires ultrashort laser pulses with with peak powers ranging from tens of TW to PW systems that are  costly and generally limited to low repetition rates (sub-\si{\Hz} to \si{\Hz}), constraining its practicality for radioisotope production. Conversely, the self-modulated regime (SM-LWFA), extensively studied in the 1990s \cite{Andreev:1992,Krall1993,Fisher1996} and later superseded by the blowout regime due to advances in laser technology, has recently regained attention with the emergence of affordable, high-repetition-rate ($\sim$kHz) TW laser systems \cite{ekspla2024, amplitude2024}. Although SM-LWFA typically produces electron beams with broader energy spreads and higher emittances \cite{Esarey2009,Maldonado2021}, it operates with longer pulses and lower peak powers (sub-\si{\tera \watt} to few \si{\tera \watt}). In this regime, laser self-focusing enhances nonlinear effects that drive self-modulation, resulting in resonant laser–plasma interactions capable of accelerating multi-MeV electron bunches. Despite the lower beam quality, the compatibility of SM-LWFA with compact, high-repetition-rate laser systems makes it an attractive alternative for applications less dependent on beam quality, such as photonuclear production of medical radioisotopes.

The highly nonlinear processes involved in the self-modulated regime of laser-driven plasma acceleration can be explored using particle-in-cell (PIC) simulations \cite{Birdsall1991}. In these simulations, the motion of charged particles is solved self-consistently under the Lorentz force, while the electromagnetic fields are computed on a spatial grid through the discretized solution of Maxwell’s equations. The PIC simulations self-consistency ensures that any acceleration mechanism---including the most common, LWFA and DLA \cite{Gahn1999} (Direct Laser Acceleration, via the Lorentz force)—are taken into account in the simulations. The phase space of the accelerated electron beam obtained from PIC simulations can be used as a particle source in Monte Carlo (MC) simulations  \cite{Agostinelli2003}, which model bremsstrahlung photon generation and their transport through matter, and subsequently estimate radioisotope production by evaluating nuclear reaction cross-sections \cite{Allison2006}. Hence, coupling PIC and MC simulations enables the systematic investigation of the influence of laser and plasma parameters on the final $^\text{99}$Mo yield.

Both PIC and MC simulations are computationally demanding, as achieving statistically and physically reliable results typically requires large particle counts, fine spatial and temporal discretization, and extensive parameter sampling. Even with GPU acceleration, a single PIC simulation can take several hours to be executed. Similarly, MC simulations that typically run on CPUs, require comparable computational time, as achieving sufficient statistical accuracy demands tracking on the order of $10^8$ particles. Consequently, performing a conventional parameter-space scan to evaluate $^\text{99}$Mo yield becomes prohibitively time-consuming.

To address this challenge, strategies that efficiently explore the parameter space, such as Bayesian optimization \cite{Brochu2010}, can be employed. By iteratively updating a probabilistic surrogate model of a given function to be maximized, i.e., the objective function, Bayesian optimization efficiently identifies promising regions of the parameter space while minimizing the number of costly simulations. The use of machine learning algorithms, particularly Bayesian optimization, to improve laser–plasma acceleration systems has been an active area of research in recent years, with numerous studies demonstrating its effectiveness in identifying optimal LWFA configurations \cite{Shalloo2020,Jalas2021,Ye2022,Irshad2023,Pousa2023,Jiang2025,Nunes2025,Valenta2025,Tchetovsky2026,Guan2026}. 

In a previous study \cite{Nunes2025} from our group, Bayesian optimization was employed to maximize the laser-to-beam energy conversion efficiency in LWFA simulations. After the optimization, the resulting energy spectrum was used, along with a simple beam spatial distribution, to evaluate the $^\text{99}$Mo yield by a MC simulation. Although this approach provided valuable insights, it did not directly optimize the isotope yield, nor did it account for the coupling between the electron-beam phase-space properties and the subsequent photon and radioisotope yields. In this work, these limitations are addressed by incorporating the MC simulation directly into the optimization loop and using  the $^\text{99}$Mo yield as the objective function to be maximized, which resulted in an order of magnitude improvement on this radioisotope yield, when compared with the previous work \cite{Nunes2025}.

\section{\label{sec:model}The Model}

The system modeled in the PIC simulations consists of a Gaussian laser pulse propagating through a transversely uniform, longitudinally varying density profile of neutral hydrogen gas. The laser peak power, pulse duration, and focal spot size (beamwaist) are constant, while the focal position along the gas target is one of the parameters to be optimized. The gas target  starts at $z=0$, and  features an asymmetric trapezoidal density profile, which is a good approximation of targets that can be experimentally obtained using asymmetric de Laval nozzles \cite{Couperus2016,Chiomento2021}, including adjustable \cite{Lei2023}, modular \cite{Lei2024}, or staged \cite{Tomkus2024} designs. 

Each segment of the trapezoidal profile plays a distinct role in the laser–plasma dynamics: the plateau density and length influence both the wakefield amplitude and the dephasing length of the LWFA process \cite{Esarey2009}. The up-ramp can enhance the energy gain by extending the dephasing length and contribute to improved collimation and reduced emittance of the accelerated electron bunches \cite{Aniculaesei2019,Yu2015}. Conversely, a down-ramp tends to trigger the injection of a larger number of electrons into the accelerating structure, increasing the bunch charge \cite{Zhang2015,Swanson2017}. In addition, the plasma density decrease reduces the phase velocity of the plasma wave \cite{Fubiani2006}, which in turn affects the dephasing length \cite{Benedetti2015}.

The MC simulations model the subsequent photonuclear stage and involve three main components: the electron source, derived from the phase-space distribution of the LWFA-accelerated beam; a cylindrical tantalum converter; and a cylindrical natural molybdenum target, both with a transverse diameter of 2.5 \si{\cm}. Efficient bremsstrahlung generation requires a converter material with a high atomic number and a small thickness to minimize photon self-absorption. However, the optimal converter thickness depends on the electron energy, as higher-energy electrons require more material to be decelerated effectively. To balance these factors while keeping a small number of inputs, the converter thickness was fixed at 4.5 \si{\mm}, a value previously shown to maximize photon yield for electron beams in the 40–50 \si{\MeV} range, consistent with spectra typically produced by SM-LWFA with similar parameters\cite{Nunes2025}. The molybdenum target thickness was set to 5 \si{\cm}, sufficient to ensure that nearly all bremsstrahlung-generated photons in the energy range relevant to the $^\text{100}$Mo($\gamma$, $n$)$^\text{99}$Mo reaction, between 8~\si{\MeV} and 20~\si{\MeV}\cite{Trknyi2018}, are absorbed. Once the optimal gas-target configuration is established, these thicknesses can be refined in subsequent optimization stages.

\subsection{\label{sec:bo}Bayesian Optimization}

The optimization loop couples PIC simulations on GPUs with MC simulations on CPUs, executed sequentially on the SDumont supercomputer cluster. The Bayesian optimization framework was implemented using the Optimas library \cite{Pousa2023,Hudson2022}, with the goal of maximizing the objective function $F_\text{obj}$, defined as the total number of $^\text{99}$Mo atoms produced in the MC simulation. At each iteration, a PIC simulation is performed for a given set of laser and plasma parameters, and the resulting phase-space of the electron beam is used as input for the MC simulation to estimate the corresponding $^\text{99}$Mo yield. Starting from 10 randomly sampled sets of input parameters, the algorithm constructs a surrogate model based on Gaussian process regression \cite{Rasmussen2008}, which provides a probabilistic prediction of $F_\text{obj}$ across the parameter space. Through successive iterations, the optimization process efficiently explores this space, refining the surrogate model and progressively converging toward the parameter set that maximizes $F_\text{obj}$. For a given set of $k$ input parameters, $x_1, \ldots, x_k$, the surrogate model estimates the mean $\mu(x_1, \ldots, x_k)$ and the standard deviation $\sigma(x_1, \ldots, x_k)$ of the objective function at untested locations, representing its expected value and associated uncertainty, respectively. These quantities are used to evaluate the acquisition function, which guides the selection of the next candidate parameters for the subsequent PIC simulation \cite{Brochu2010}. In this study, the \texttt{AxSingleFidelity} generator of Optimas was employed, which internally uses the Noisy Expected Improvement acquisition function \cite{Ament2023} for single-objective optimization. This coupled and self-consistent simulation--optimization strategy enables the efficient identification of configurations that directly maximize the $^\text{99}$Mo yield.

The PIC simulation input parameters to be optimized, listed in Table I, are the laser focal position, plateau gas density, plateau length, and up- and down-ramp lengths. Together, these three lengths define a trapezoidal gas-density profile that is generally asymmetric. The output quantities (electron energy spectrum, beam emittances, divergences, and laser-to-beam energy conversion efficiency \cite{Streeter2022, Irshad2024, Nunes2025}) are obtained from the accelerated electron beam computed in each PIC simulation. In contrast to the previous approach \cite{Nunes2025}, which assigned Gaussian spatial distributions to the accelerated beam, the present framework employed the phase-space  obtained from the PIC simulations as input in the MC simulations.\color{black}

     


\begin{table}[!b]
\caption{\label{tab:inputs}PIC Input parameters.}
\centering
\begin{tabular}{@{}cccl}
\hline
\multirow{2}{*}{\makecell[c]{Parameter}} & 
\multirow{2}{*}{\makecell[c]{Range}} & 
\multirow{2}{*}{\makecell[c]{Unit}} &
\multirow{2}{*}{\makecell[c]{Description}}\\
&&\\
\hline\\[-2.2ex]

$n_\mathsmaller{\si{H}}$ & $0.1 \;,\, 10\phantom{00}$ & $10^{19}\,\si{\per\cm\cubed}$ & \makecell[l]{plateau H atom density}\\[1ex]

$R_1$ & $\phantom{0}5 \;,\, 1000$ & $\si{\um}$ & up-ramp length\\[1ex]

$L_1$ & $\phantom{0}5 \;,\, 1000$ & $\si{\um}$ & plateau length\\[1ex]

$R_2$ & $\phantom{0}5 \;,\, 1000$ & $\si{\um}$ & down-ramp length\\[1ex]

$z_\mathrm{foc}$ & $\phantom{0}5 \;,\, 1500$ & $\si{\um}$ & laser focal position\\
\hline
\end{tabular}
\end{table}

\subsection{\label{sec:pic}Particle-in-cell Simulations}

PIC simulations were conducted using the Fourier-Bessel PIC (FBPIC) code \cite{Lehe2016} with eight GPUs NVIDIA V100 (32 GB). FBPIC uses a spectral solver on 2D (RZ) grids, each representing an azimuthal mode $m$. Furthermore, the spectral solver algorithm reduces spurious numerical dispersion, including the zero-order numerical Cherenkov effect \cite{Godfrey1974}. 

Quasi-cylindrical pseudo-spectral PIC codes, by incorporating a finite number of azimuthal modes, can capture the essential physical dynamics while accommodating moderate asymmetries, providing a balance between computational efficiency and accuracy compared to fully 3D simulations. 
In this study, three azimuthal modes were employed, consistent with the FBPIC documentation’s recommendation for modeling nonlinear phenomena such as electron self-injection. Nonetheless, fully Cartesian 3D simulations may still be required in strongly asymmetric scenarios or when a high-fidelity representation of complex transverse dynamics is necessary.

In all simulations, a linearly $x$-polarized laser pulse with a peak power of 10~\si{\tera\watt} propagates along the $z$ direction through a neutral hydrogen density profile. The pulse is modeled as a Gaussian beam, focused to a waist of $w_0 = 7$~\si{\um} at the focal position $z_{foc}$. During propagation, the local plasma electron density is computed using the ADK ionization model \cite{Ammosov1986}. These values are consistent with previous studies \cite{Salehi2017, Maldonado2021, Goers2015, Woodbury2018} and are compatible with the design of a self-modulated LWFA currently under development at the Center for Lasers and Applications in the Nuclear and Energy Research Institute (IPEN-CNEN, Brazil) \cite{Bonatto:2021}. The laser parameters and numerical simulation settings, such as domain size, spatial resolution, number of azimuthal modes, and particles per cell, are listed in Table~\ref{tab:laser_PIC}. A fixed random seed was used throughout to ensure reproducibility and to reduce stochastic variability during the optimization process.

\begin{table}[!t]
\caption{\label{tab:laser_PIC} Laser parameters and PIC settings.}
\centering
\begin{tabular}{@{}cccl}\hline

\multirow{2}{*}{\makecell[c]{Parameter}} & \multirow{2}{*}{\makecell[c]{Value}} & \multirow{2}{*}{\makecell[c]{Unit}} &\multirow{2}{*}{\makecell[c]{Description}}\\
&&\\
\hline\\[-2.2ex]

$P_L$ &  10 & \si{\tera\watt} & initial peak power\\[1.0ex]
$\lambda_0$ & 800 & \si{\nm} & wavelength  \\[1.0ex]
$z_0$ & -25 & \si{\um} & pulse centroid\\[1.0ex]
 $\tau$ & 35 & \si{\femto s} & intensity pulse duration (FWHM)\\[1.0ex]
$w_0$ & 7 & \si{\um} & beam waist\\[1.0ex]
$a_0$ &  2.47 & --- & strength parameter  \\[1.0ex]
$E_\ell$ & 372 & \si{\mJ} & pulse energy\\[1.0ex]
  \multirow{2}{*}{$z_\mathrm{min}$} &  \multirow{2}{*}{-75} & \multirow{2}{*}{\si{\um}} &
 \multirow{2}{*}{\makecell[l]{initial boundary of the longitudinal\\simulation domain}} \\
  &  & & \\[1.0ex]
 \multirow{2}{*}{$z_\mathrm{max}$} & \multirow{2}{*}{0} & \multirow{2}{*}{\si{\um}} &  \multirow{2}{*}{\makecell[l]{final boundary of the longitudinal\\simulation domain}}\\
  &&&\\[1.0ex]
  \multirow{1}{*}{$r_{\mathrm{max}}$} &  \multirow{1}{*}{65} &   \multirow{1}{*}{\si{\um}} &  \multirow{1}{*}{\makecell[l]{ radius of the simulation edge}}\\[1.0ex]
 $\Delta_z$ & 27 & \si{\nm} & spatial resolution in $z$  \\[1.0ex]
 $\Delta_r$ & 33 & \si{\nm} & spatial resolution in $r$ \\[1.0ex]
   
 $N_m$ & 3 & --- & number of azimuthal modes  \\[1.0ex]
 $N_\mathrm{pz}$ & 2 & --- & particles per cell along $z$  \\[1.0ex]
 $N_\mathrm{pr}$ & 2 & --- & particles per cell along $r$  \\[1.0ex]
 $N_\mathrm{p\theta}$ & 12 & --- & particles per cell along $\theta$  \\        
 \hline
%
\end{tabular}

\end{table}

\subsection{\label{sec:mc}Monte Carlo Simulations}

Monte Carlo simulations were performed with the openTOPAS code \cite{Faddegon2020,Perl2012}, using as input the electron beam phase space obtained from the FBPIC simulations. Since the PIC simulation runs on GPUs while the MC simulation runs on CPUs, the overall efficiency of the optimization loop depends on minimizing GPU idle time. A pilot study showed that limiting the number of primary electrons in the MC simulations to $N_e = 10^7$ provides a good cost-benefit ratio: the electron spectra are accurately reproduced with acceptable statistical uncertainty while maintaining a sufficiently short CPU runtime.

The electron bunch generated by FBPIC is represented by $N$ macroparticles, each carrying a weight $w_j$ corresponding to the number of physical electrons it represents. In general, the number of physical particles in the accelerated FBPIC beam is much larger than the number of MC histories adopted in openTOPAS ($N_e=10^7$), making it necessary to sample a representative subset while preserving the statistical properties of the original beam. Because the cross section of the $^\text{100}$Mo($\gamma$, $n$)$^\text{99}$Mo reaction is significant only for photons in the 8--20~\si{\MeV} range, only electrons with kinetic energies above 8~\si{\MeV} were considered in the sampling procedure. The total charge associated with these selected electrons is denoted by $Q_{\mathrm{sel}}$. 

From this selected energy subset, $N_e$ macroparticles were sampled according to the discrete probability distribution
\begin{equation}
p_j = \frac{w_j}{\sum_{i}^N w_i},
\end{equation}
where $p_j$ is the probability of selecting the macroparticle with index $j$. The phase-space coordinates (positions and momenta) of the sampled macroparticles were then written into the openTOPAS input file, with each sampled macroparticle initiating one MC history.

To scale the MC results to the physical beam charge, each history was assigned a weight $W_{\mathrm{MC}} = Q_{\mathrm{sel}}/(e N_e)$, where $e$ is the elementary charge. All scored quantities were multiplied by $W_{\mathrm{MC}}$ to obtain estimates corresponding to the actual number of electrons in the beam.

In the MC simulations, the electron beam propagates through  10~\si{\cm} of vacuum before striking a Ta converter placed immediately before a natural molybdenum target. The physics modules \texttt{g4em-standard\_opt4} and \texttt{g4em-extra} were employed, and the $^\text{99}$Mo yield was quantified using a \texttt{Scorer(OriginCount)} detector.

Although a recent study \cite{Nunes2026} from our group showed that a multigroup method can estimate $^\text{99}$Mo production in a fraction of a second and could replace MC simulation in the optimization process, it relies on a simplified beam model that considers only the electron energy spectrum. In contrast, using the full PIC-derived phase-space distribution provides a more realistic description of the electron beam and its impact on photon and radioisotope production.

\color{black}

\section{\label{sec:results1} System optimization}

\begin{figure}[!t]
    \centering
    \includegraphics{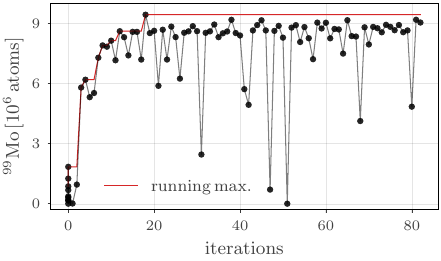}
    \caption{Number of $^\text{99}$Mo atoms produced at each iteration (black dots). The red line indicates the highest value obtained in the optimization up to that point.} 
    \label{fig:caso1_iter}
\end{figure}

Figure~\ref{fig:caso1_iter} shows that after 10 iterations, the objective function already converged to approximately $8.6 \times 10^6$ atoms of $^\text{99}$Mo, around which it continued to oscillate throughout the remaining iterations, aside from a few ``outliers'' corresponding to exploratory probes of low-yield regions of the parameter space. The maximum value, $9.4 \times 10^6$ atoms of $^\text{99}$Mo, was obtained at iteration 18. As no further improvement was observed in the remaining iterations, the optimization was stopped. The ``outliers'', which deviate from the optimal value, reflect the exploratory behavior of the algorithm, which occasionally tests new combinations of input parameters in lower-yield regions of the parameter space to avoid an overly restrictive search around the current optimum, and to reduce the likelihood of premature convergence to local optima.

\begin{figure*}[!t]
    \centering
    \includegraphics[scale=0.98]{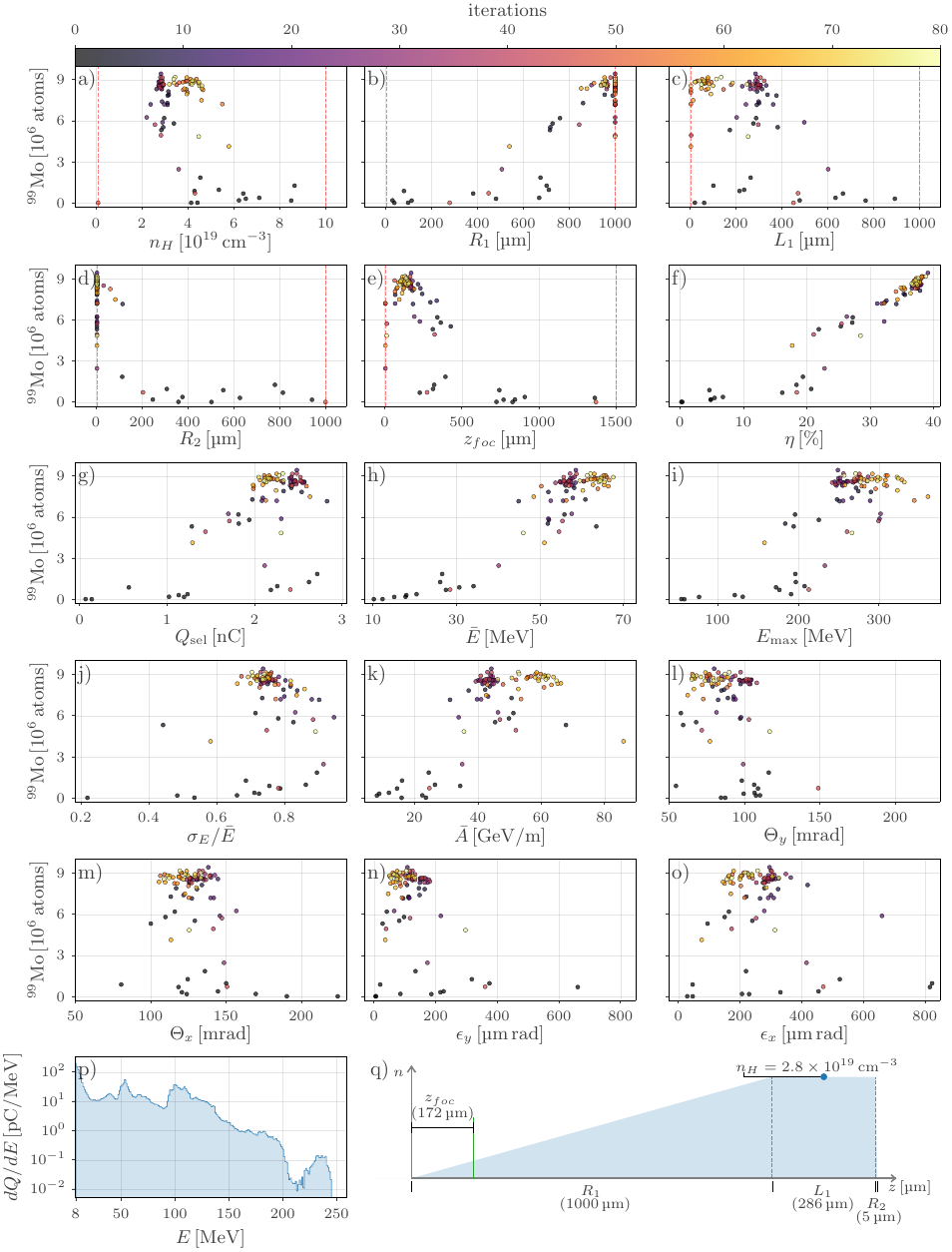}
    \caption{Results of the Bayesian optimization. In the scatter plots, the color scale indicates the algorithm iteration. Input parameters: $^\text{99}$Mo yield as a function of (a) plateau hydrogen density, (b) up-ramp length, (c) plateau length, (d) down-ramp length of the gas-density profile, and (e) laser focal position. Output parameters: $^\text{99}$Mo yield as a function of (f) laser-to-beam energy conversion efficiency, (g) charge of particles with final energies above 8~\si{\MeV}, (h) average energy, (i) maximum energy, (j) energy dispersion, (k) average acceleration gradient, divergences along the (l) $y$ and (m) $x$ directions, and normalized emittances along the (n) $y$ and (o) $x$ directions. Optimal simulation: (p) electron energy spectrum and (q) gas-density profile, showing the optimal input-parameter values.}
    \label{results2}
\end{figure*}

Figures~\ref{results2}(a)–(e) present scatter plots illustrating the evolution of the objective function as a function of each input parameter listed in Table~\ref{tab:inputs} during the Bayesian optimization process. In each panel, the red dashed lines indicate the bounds of the parameter space, while the color scale encodes the iteration number, progressing from dark (early iterations) to bright (later iterations). Figure~\ref{results2}(a) indicates that the optimization converged toward systems with plateau densities in the range $(2 < n_\mathsmaller{\si{H}} \lesssim 5)\times10^{19}$ \si{\per\cm\cubed}, reaching a maximum at $2.84\times10^{19}$ \si{\per\cm\cubed}. In later iterations, the algorithm also explored configurations with higher densities, around $4\times10^{19}$ \si{\per\cm\cubed}, which resulted in $^\text{99}$Mo production comparable to that at the peak. Regarding the up-ramp length $R_1$, Fig.~\ref{results2}(b) indicates that the best results were obtained for the maximum allowed value, $1000$ \si{\um}. Although longer ramps would likely further increase the $^\text{99}$Mo production, the parameter range was not extended due to computational cost constraints. 

Fig.~\ref{results2}(c) shows that the optimization converged toward a plateau length near $300$ \si{\um}, with the best result obtained for $L_1 = 286$ \si{\um}. Nevertheless, the cluster of yellow-orange points with lengths under 200 µm indicates that shorter plateaus can also yield comparable results. These correspond to the same yellow points in Fig.~\ref{results2}(a), associated with $n_\mathsmaller{\si{H}} \approx 4\times10^{19}$ \si{\per\cm\cubed}, suggesting that a shorter plateau can be compensated by a higher density while maintaining a similar $^\text{99}$Mo yield, thereby revealing an additional local maximum of the system. Figure~\ref{results2}(d) shows that the down-ramp length converged to the minimum allowed value, $R_2 = 5$ \si{\um}, corresponding to an extremely abrupt drop in plasma density. As will be discussed later, this behavior suggests that, once electrons have gained energy from the wakefield, it is advantageous for them  that the plasma ends abruptly, preventing these particles from entering the decelerating phase of the wakefield. Regarding the laser focal position, Fig.~\ref{results2}(e) shows that the optimization converged to the range $(50 \lesssim z_{\mathrm{foc}} \lesssim 250)$ \si{\um}, with the best result obtained at $z_{\mathrm{foc}} \approx 172$ \si{\um}, near the beginning of the up-ramp. The input parameter $z_{\mathrm{foc}}$ is defined for laser propagation in vacuum; once the laser interacts with the plasma, effects such as relativistic self-focusing \cite{Esarey2009} modify the beam evolution, shifting the effective focal position and extending the focal region.

Figures~\ref{results2}(f)–(o) show scatter plots of the output parameters, which characterize the properties of the accelerated electron bunch in each simulation. Figure~\ref{results2}(f) presents the laser-to-beam energy conversion efficiency \cite{Streeter2022, Irshad2024, Nunes2025}, $\eta$, considering only particles with energies above  8 \si{\MeV}. During the optimization, efficiencies exceeding 30\% were achieved, with the best simulation reaching 39.1\%, representing a significant improvement over values previously reported in the literature, which typically range from 5\% to 13\% \cite{Irshad2024, Gotzfried2020, Feng2023, Nunes2025}. A clear correlation is observed between $\eta$ and the number of $^\text{99}$Mo atoms produced: as a larger fraction of the laser energy is transferred to the electron beam, more high-energy electrons become available to generate photons capable of inducing the $^\text{100}$Mo($\gamma$,n)$^\text{99}$Mo reaction.

Figure~\ref{results2}(g) shows the charge of the electron beams. Although $^\text{99}$Mo yields above $6\times10^6$ atoms were obtained for beam charges between 1.7 and 3 \si{\nano\coulomb}, a few simulations within the 2-3 \si{\nano\coulomb} range produced fewer than $3\times10^6$ $^\text{99}$Mo atoms, indicating that beam charge alone does not determine the yield. \color{black}As shown in Fig.~\ref{results2}(h), the electron energy also plays a crucial role: these lower-yield beams had average energies below 41 \si{\MeV}, whereas the best results were achieved for average energies above 45 \si{\MeV}. Regarding the maximum beam energy, Fig.~\ref{results2}(i) shows that the best-performing cases include electrons with energies between 250 and 300 \si{\MeV}, although only a small fraction of particles reach these values due to the exponential decay of the charge distribution. This behavior is consistent with the results of our previous work \cite{Nunes2026}, which showed that high energy electrons are more efficient than less energetic ones when producing  $^\text{99}$Mo atoms; this is also backed up by the energy spread shown in Fig.~\ref{results2}(j), where a cluster of points between 0.7 and 0.8 indicates a broad energy dispersion.

Figure~\ref{results2}(k) presents the average accelerating gradient estimated from the target length (given by its end position), which exhibits two well-separated clusters: one composed of darker (purple) points around 45 \si{\MeV}, and another of lighter points near 60 \si{\MeV}. This behavior arises from the higher average energies of these beams, as shown in Fig.~\ref{results2}(h), combined with shorter plateau lengths. As indicated in Fig.~\ref{results2}(c), these lighter-colored points correspond to profiles with approximately half the plateau length of those associated with the darker points.

Figures~\ref{results2}(l–o) present beam quality parameters, namely divergence and emittance. Although these quantities are not critical for $^\text{99}$Mo production—since photon generation primarily depends on the electron–target interaction—it is noteworthy that the optimization process naturally converges toward configurations that partially minimize both divergence and emittance, resulting in a more collimated beam. This occurs even though these parameters were not explicitly included in the objective function. Additionally, this information may be relevant for other SM-LWFA applications in which beam quality is important. Fig.~\ref{results2}(l) and (m) show the beam divergence in the $y$ and $x$ directions, respectively, and Fig.~\ref{results2}(n) and (o) present the corresponding RMS emittances. While in the $y$ direction the best-performing cases exhibit divergences between 50 and 100 \si{\milli\radian}, the divergence in the $x$ direction, aligned with the laser polarization, is systematically higher by about 50 \si{\milli\radian}. A similar trend is observed for the emittance: in the $y$ direction, there is a cluster of points between 50 and 200 \si{\um\radian}, whereas in the polarization direction ($x$) the emittance increases by nearly 200 \si{\um\radian}.

Figures~\ref{results2} (p) and (q) present the optimal beam energy spectrum and the corresponding gas-density profile, respectively. Table~\ref{tab:optimal_results} summarizes the input and output parameters for the optimal simulation.

\begin{table}[!t]
    \caption{\label{tab:optimal_results} Optimal simulation input and output parameters.}
    \centering
    \begin{tabular}{ccc}\hline
        \multirow{2}{*}{Parameter} & \multirow{2}{*}{Value} & \multirow{2}{*}{Unit} \\
        &&\\
        \hline\\[-2.2ex]
        $n_\mathsmaller{\si{H}}$ & 2.84 & $10^{19}\,\si{\per\cm\cubed}$ \\[0.70ex]
        $R_1$                    & 1000  & \si{\um}    	                 \\[0.70ex]
        $L_1$                    & 286.5 & \si{\um}                      \\[0.70ex]
        $R_2$                    & 5     & \si{\um}                      \\[0.70ex]
        $z_\mathrm{foc}$                & 171.9 & \si{\um}	                     \\[0.70ex]
        \hline\\[-2.2ex]
        $^\text{99}$Mo                & 9.42 & $10^6$ \si{atoms}         \\[0.70ex]
        $\eta$                   & 39.1             & \si{{\%}}          \\[0.70ex]
        $Q_\mathrm{sel}$                & 2.48             & \si{\nano\coulomb}  \\[0.70ex]
        $\bar{E}$                & 58.6             & \si{\MeV}          \\[0.70ex]
        $E_{\mathrm{max}}$         & 249              & \si{\MeV}          \\[0.70ex]
        $\sigma_{E} / \bar{E}$   & 0.74             & ---  \\[0.70ex]
        $\bar{A}$                & 44.3             & \si{\giga\eV/\m}   \\[0.70ex]
        $\Theta_y$               & 80               & \si{\milli\radian} \\ [0.70ex]
        $\Theta_x$               & 138              & \si{\milli\radian} \\ [0.70ex]
        $\epsilon_y$             & 109               & \si{\um \radian} \\ [0.70ex]
        $\epsilon_x$             & 296              & \si{\um \radian} \\ [0.70ex]
        \hline
    \end{tabular}
\end{table}



\subsection{\label{sec:dynamics}Optimal Simulation Dynamics}

\begin{figure*}[!t]
    \centering
    \includegraphics{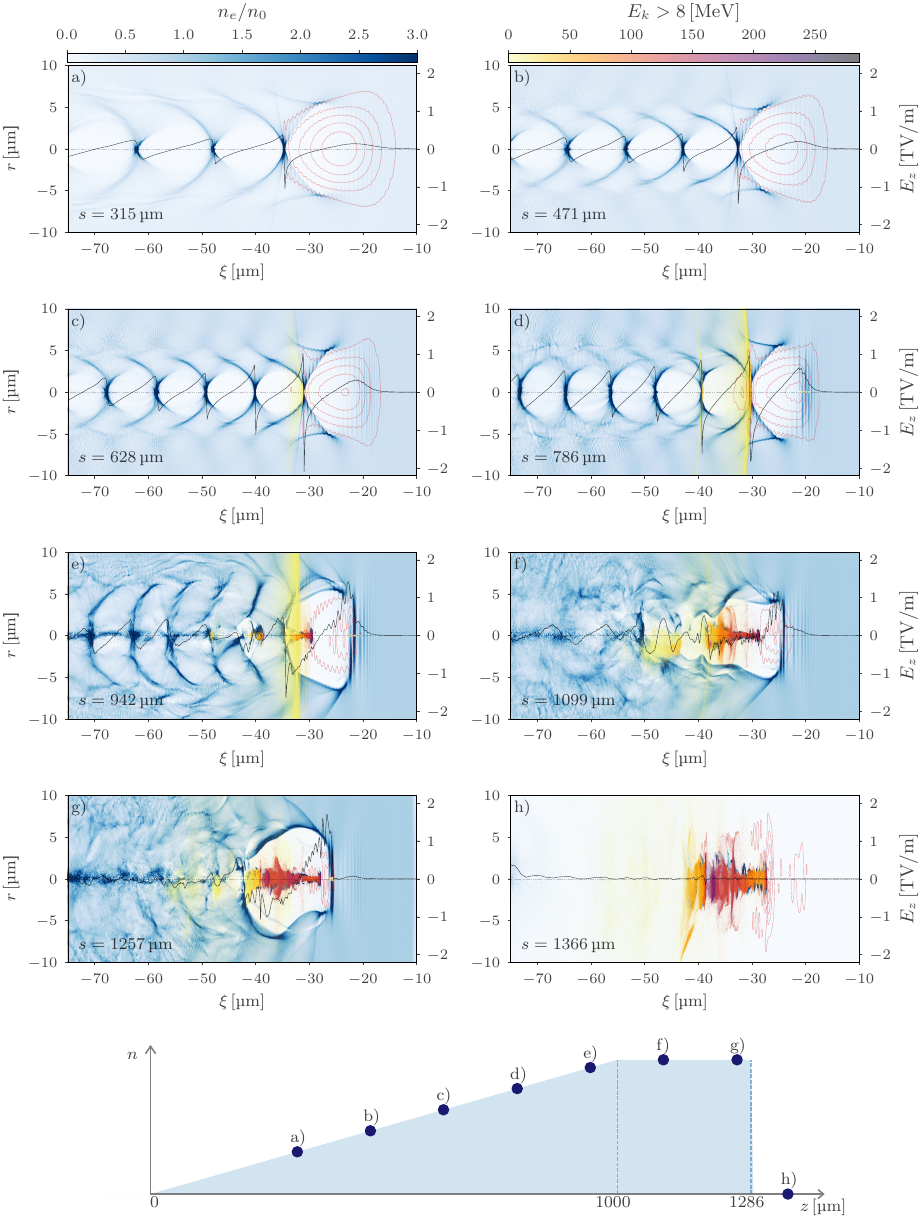}
    \caption{Spatial evolution of the optimal simulation. Panels (a)–(h) display the plasma electron density, normalized by the plateau density, in a blue gradient scale. The backtracked electrons—those that later reach energies above 8 \si{MeV} after ejection—are shown as colored dots, where the yellow–black color scale represents their instantaneous energy at each propagation distance. Contour lines of $a_0$ indicate the laser envelope at each position, and the black curve denotes the longitudinal electron field $E_z$ along the plasma axis. The lower panel shows the target profile and the positions for which panels (a)–(h) were plotted.} 
    \label{nelec2}
\end{figure*}

The underlying physics of the optimal simulation can be elucidated by backtracking the trajectories and energy evolution of the electrons in the accelerated beam, composed by those exiting the gaseous target with energies above 8~\si{MeV}. Figures~\ref{nelec2}(a)–(h) present the electron density, normalized by the plateau density ($n_e/n_\mathsmaller{\si{H}}$), displayed in a blue color scale. The backtracked electrons are overlaid as colored dots, with the yellow-to-black color scale indicating their instantaneous energies. Superimposed red contour lines represent the laser envelope normalized by the laser strength parameter $a_0$. The contour levels are automatically scaled according to the local maximum laser envelope value, resulting in slightly different ranges between panels (e.g., from $0 - 3\,a_0$ at early stages to $0 - 7.5\,a_0$ at later ones). The outermost contours denote lower field amplitudes, while the innermost correspond to higher laser intensities. The right vertical axis corresponds to the on-axis longitudinal electric field $E_z$, represented by the superimposed black line along the propagation axis—that is, the center of the plasma. Each panel in this figure is plotted at a distinct position $s=v_w t$ (indicated at its bottom-left corner), which is the origin of the coordinate $\xi=z-s$; $v_w$ is the velocity of the moving window, chosen to  follow the laser pulse, as done in a previous study~\cite{Nunes2022}. The figure bottom (unlabeled) panel shows the  position $s$ at which each panel is plotted.


Figure~\ref{nelec2}(a) shows a focused laser pulse located approximately at one-third of the target up-ramp. Within the pulse, the intense electromagnetic fields cause a depletion of the normalized plasma electron density ($n_e/n_\mathsmaller{\si{H}}$), visible as a white region, ahead of bubble-like cavities. The on-axis longitudinal electric field inside these structures exhibits a well-defined profile, evolving toward an almost sawtooth-like waveform in the last bubbles and reaching nearly $-1$~\si{\tera\volt/\meter} at the back of the first one. At this position, two symmetrical diagonal light-blue lines originate from $r=0$~\si{\um}, indicating electron ejection likely caused by Coulomb repulsion due to the high electron density behind the first bubble, combined with a defocusing (negative) transverse wakefield. 

As the laser propagates along the up-ramp, the rising plasma density shortens the wakefield wavelength, so that more bubbles fit behind the driver. Figure ~\ref{nelec2}(b) shows this regime near the midpoint of the up-ramp, where five distinct bubbles are visible within the simulation domain. At this point, the minimum $E_z$ value reaches approximately $-1.5$~\si{\tera\volt/\meter} at the back of the first bubble, while the field in the second one becomes less linear. Fig.~\ref{nelec2}(c) shows that at $z\approx600\,\si{\um}$, the first electrons that will form the beam (those leaving the plasma with energies above 8 MeV) begin to appear as light yellow dots at the rear of the first bubble. Despite being transversely ejected at this point, these electrons are recaptured downstream and form the final beam. At this stage, $E_z$ reaches its overall minimum value of approximately $-2$~\si{\tera\volt/\meter}. Figure~\ref{nelec2}(d) reveals a slight modulation in the laser pulse, with a small fragment appearing at the front of the second bubble, as indicated by the laser contour lines. Additionally, the $E_z$ values in the second bubble become comparable to those in the first one, while yellow dots show that particles are ejected at the back of the first two bubbles. 

As the laser approaches the end of the up-ramp, it becomes narrower [Fig.~\ref{nelec2}(e)], and self-injection occurs at the rear of the pulse. The injected electrons are pushed forward, being accelerated up to nearly 200~\si{\MeV} in the first bubble ($\xi \approx -30$~\si{\um}), while a second, lower-energy bunch forms in the middle of the second bubble ($\xi \approx -40$~\si{\um}). At this stage, the plasma dynamics become increasingly complex: the first bubble lengthens, low-energy electrons from the trailing bubbles are accelerated toward the leading ones, and the bubble structures at the back of the window begin to merge and deteriorate. When the system reaches $z \approx 1100$~\si{\um} (at the plasma plateau), the trailing bubbles completely disappear, as shown in Fig.~\ref{nelec2}(f). Two distinct electron bunches are now visible: a lower-energy one, between 
$-50 < \xi < -41$~\si{\um}, and a higher-energy one, at $\xi > -40$~\si{\um}, with a few electrons exceeding 250~\si{\MeV}. Figure~\ref{nelec2}(g) shows the plasma near its end, where a high-charge, high-energy beam occupies the first bubble. The beam front enters the decelerating phase and starts losing energy, while the rear continues to gain it. The target then abruptly terminates, producing a sudden drop in density; as a result, the accelerated beam inside the bubble structure remains nearly unchanged, preserving its properties as it exits the target, as shown in Fig.~\ref{nelec2}(h). Once outside the gas target, the beam expands transversely due to the absence of focusing forces and the electrons Coulomb repulsion.

Figure~\ref{nelec2} suggests that the dominant acceleration mechanism in the optimal simulation is LWFA, as indicated by the small number of electrons located within the laser contour lines. This can be verified by analyzing the relative contributions of LWFA and DLA (which occurs when the electrons reach relativistic speeds and are pushed forward by the Lorentz force within the laser pulse) to the electrons kinetic energy gain  \cite{Cohen2024, Nunes2025}. The kinetic energy of an electron is given by $E_k = m_ec^2(\gamma-1)$, where $\gamma^2 = \Gamma_{\parallel} + \Gamma_{\perp} + 1$. In this expression, $\Gamma_{\parallel}$ represents the normalized longitudinal energy gain associated with the wakefield (LWFA), while $\Gamma_{\perp}$ denotes the normalized transverse energy gain arising from the interaction with the laser’s transverse electric field (DLA). 

\begin{figure*}[!t]
\centering
\includegraphics[scale=1]{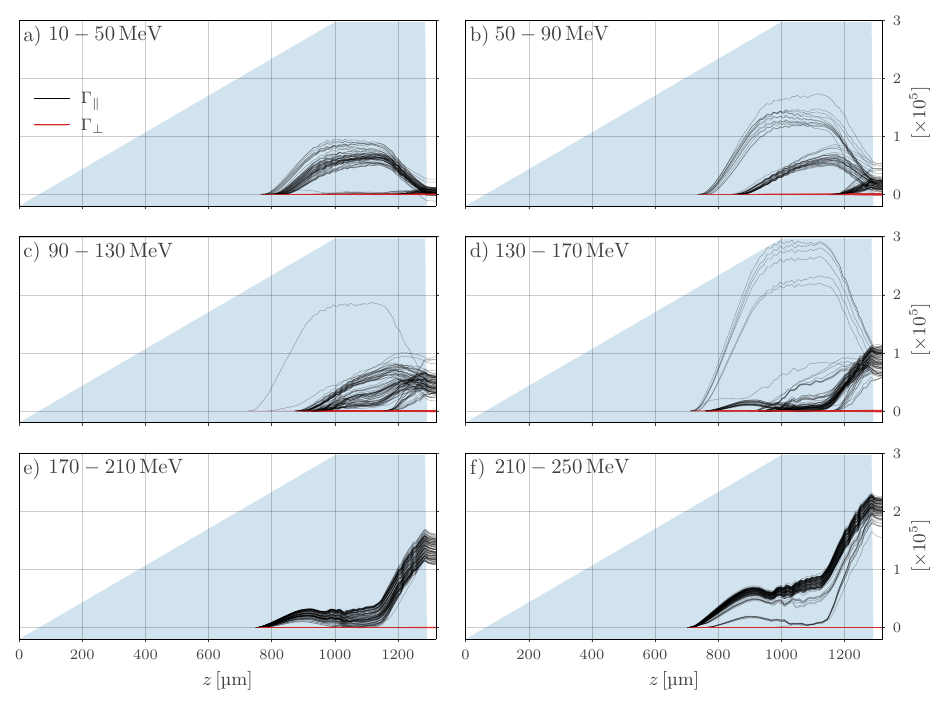 }
\caption{Contributions of LWFA and DLA to the total electron energy gain in the optimal simulation.}
\label{fig:lwfavsdla}
\end{figure*}

Figures~\ref{fig:lwfavsdla}(a–f) present the LWFA ($\Gamma_\parallel$) and DLA ($\Gamma_\perp$) energy gain contributions, shown as black and red lines, respectively, for the optimal simulation. Each panel displays 100 electrons, evenly distributed over a 40~\si{\MeV} interval in the final electron energy. Figure~\ref{fig:lwfavsdla} indicates that LWFA is the dominant---and effectively the only---acceleration mechanism, regardless of the electrons’ final energies. Since $\Gamma_\perp \approx 0$ for all energy ranges, $\gamma^2 \approx \Gamma_\parallel$. However, an analysis of the electron evolution throughout the LWFA acceleration dynamics provides a deeper understanding of the underlying physics.

In the $10$--$50$~\si{\MeV} range [Fig.~\ref{fig:lwfavsdla}(a)], all electrons are self-injected between $z = 780$~\si{\um} and $850$~\si{\um} and gain LWFA energy until the bubble midpoint. A fraction of these electrons have a fast energy increase, reaching the  midpoint quickly and oscillating around the accelerating–decelerating transition before abruptly losing almost all their energy (trapezoidal profile). The remaining electrons gain energy gradually and stay in the accelerating phase longer, reaching the midpoint later, and lose energy in the decelerating phase, as the former electrons. 

For the $50$--$130$~\si{\MeV} interval [Figs.~\ref{fig:lwfavsdla}(b) and (c)], three behaviours appear: trapezoidal gain, slow acceleration with late loss, and electrons injected near the target end ($z \approx 1200$~\si{\um}) that accelerate almost linearly over a short distance, and retain all acquired energy.

Panel (d) ($130$--$170$~\si{\MeV}) shows four behaviours: trapezoidal gain (injection at $z \approx 700$~\si{\um}); two groups injected at $z \approx 900$~\si{\um} and $1150$~\si{\um} that gain energy and keep it, the latter with a stronger gradient; and the fourth type---also in panels (e) and (f)---injected at $z \approx 700$~\si{\um}, losing nearly all energy by $z \approx 900$~\si{\um}, staying on a low-energy plateau, and being rapidly reaccelerated after $z \approx 1150$~\si{\um}.

In the highest-energy panels [Figs.~\ref{fig:lwfavsdla}(e) and (f)], after strong deceleration near $z \approx 900$~\si{\um}, some electrons lose almost all energy while others keep a significant fraction, affecting their final energy. In Fig.~\ref{fig:lwfavsdla}(f) the loss is  smaller; instead of remaining constant, the electrons slowly gain energy before a final rapid acceleration. A small subset injected at $z \approx 750$~\si{\um} (rather than $700$~\si{\um}) has a smaller initial gain, and after losing most of the energy,  experiences a stronger gradient, catching up with the other electrons.

The average accelerating gradient, estimated from the total plasma length and the average final energy, is 44~\si{\GeV/\m} (Table \ref{tab:optimal_results}). However, Fig.~\ref{fig:lwfavsdla} shows that most electrons composing the accelerated beam actually begin accelerating at $z \approx 700$~\si{\um}. Considering the propagation distance from this point and the average final energy in each panel, the resulting average gradients increase substantially, reaching 51, 118, 186, 254, 321, and 389~\si{\GeV/\m} for Figs.~\ref{fig:lwfavsdla}(a–f), respectively.

\subsection{\label{sec:activity} Estimation of $^\text{99}$Mo and $^{99\text{m}}$Tc Activities}

\begin{figure}[!t]
\centering
\includegraphics[scale=1]{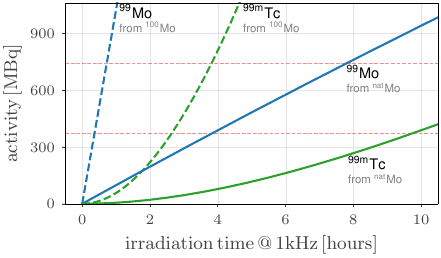}
\caption{$^\text{99}$Mo and $^{99\mathrm{m}}$Tc activities as a function of irradiation time, assuming a laser repetition rate of 1~\si{\kHz}. Solid lines represent the activities expected for a natural Mo target, whereas dashed lines correspond to those estimated for a fully enriched $^\text{100}$Mo target.}
\label{fig:activity}
\end{figure}

The administered activity of a radiopharmaceutical depends on the specific medical procedure. For $^{99\text{m}}$Tc, a typical activity of 370~\si{\mega\becquerel} per patient is commonly adopted, while US standards recommend values up to 740~\si{\mega\becquerel} \cite{Vieira2021,Drozdovitch2015}. Fig.~\ref{fig:activity} shows the activities of $^\text{99}$Mo and $^{99\text{m}}$Tc as a function of the irradiation time for the optimal simulation (Table \ref{tab:optimal_results}), considering a 1~\si{\kHz} repetition rate for the laser, for $^\text{nat}$Mo and fully enriched $^\text{100}$Mo. The $^\text{99}$Mo yield per laser shot when using the $^\text{100}$Mo target was considered to be 10.4 times that of the natural molybdenum ($9.8\times 10^7$ atoms/shot). The time evolution of the nuclear populations follows the equations for a two-member decay chain with continuous production of the parent isotope\cite{Eisberg1985}:
\begin{align*}
\frac{dN_1}{dt} &= Yf-\lambda_1 N_{1},\\
\frac{dN_2}{dt} &= \lambda_1 N_1
-\lambda_2 N_2,
\end{align*}
where $Y$ is the number of $^\text{99}$Mo nuclei produced per laser shot, $f$ is the laser repetition rate, $N_1$ and $N_2$ are the numbers of parent ($^\text{99}$Mo) and daughter ($^\text{99m}$Tc) nuclei, and $\lambda_1,\, \lambda_2$ are their respective decay constants, given by
$\lambda_i=\ln(2)/T_{1/2,i}$, where $T_{1/2,i}$ is the half-life of the $i$-th isotope. The activities are then obtained as $A_i(t)=\lambda_iN_i(t)$. The previous equations are an approximation that do not consider the $^\text{99}$Mo and $^\text{99m}$Tc decay between laser pulses\cite{Roberts2015}, but can be used due to the pulses short period (1 \si{\milli\s}) when  compared to the radioisotopes half-lives (hours).

Figure~\ref{fig:activity} shows that the optimized configuration substantially increases the activity production rate compared with the best case reported in our previous study~\cite{Nunes2025}. For the natural Mo target,   the $^{99\text{m}}$Tc clinical activity of $370~\si{\mega\becquerel}$ is obtained after approximately 10~\si{\hour} of continuous irradiation; using the fully enriched $^\text{100}$Mo target, with  the yield increased by a factor 10, reduces the irradiation time required to reach the clinical activity  to less  3 hours. Although further development and validation are required for routine clinical deployment, these results indicate a clear pathway toward the practical implementation of this production approach.

The improvement of more than one order of magnitude over the results reported in the previous study~\cite{Nunes2025} reflects the combined effect of higher laser power and a longer acceleration length in the present setup. This improvement also stems from the Bayesian optimization loop encompassing the entire system, from electron acceleration to the production of $^\text{99}$Mo. Rather than optimizing only the electron beam, the algorithm selects the beam characteristics that maximize the yield of this radioisotope through the $^\text{100}$Mo($\gamma$,n)$^\text{99}$Mo reaction, representing a clear advancement over our previous work.

\color{black}


\section{\label{sec:conclusions}Conclusions and discussion}

In this work, Bayesian optimization was applied to tandem PIC and MC simulations to maximize $^\text{99}$Mo production via the $^\text{100}$Mo($\gamma$, $n$)$^\text{99}$Mo reaction in a tantalum-molybdenum target, triggered by electrons accelerated in SM-LWFA using a 10~\si{\tera\watt} laser. While the algorithm effectively maximized the production of $^\text{99}$Mo for one order of magnitude over previous studies, other unwanted photonuclear reactions---which could generate neutrons and secondary particles---were not considered with the optimization loop \cite{Martin2017}. 

In the optimal PIC simulation, a beam with an average energy of 58.6~\si{\MeV} and a charge of 2.48~\si{\nano\coulomb}—corresponding to a laser-to-beam energy conversion efficiency of 39\%—was obtained. Monte Carlo simulations show that this beam could produce $9.42\times10^6$ atoms of $^\text{99}$Mo per laser shot. This yield, combined with kHz repetition rate lasers, could enable the production of clinical activities of $^\text{99m}$Tc within a few hours, opening the possibility of decentralized isotope production at healthcare centers, although more studies are needed.

For the optimal gas-density profile, the laser–plasma dynamics along the ramps and plateau was analyzed and discussed. Additionally, the electrons comprising the final accelerated beams were backtracked to assess their energy gains throughout propagation. Although both acceleration mechanisms—LWFA and DLA—can coexist in the self-modulated regime, the Bayesian optimization algorithm converged to a configuration in which only LWFA contributed. Average accelerating gradients between 51 and 389~\si{\giga\eV/\m} were estimated for the particles composing the accelerated beam.

The optimization yielded an abrupt density profile with a down-ramp of only 5~\si{\um}, which may be challenging to implement experimentally. However, tailored gas-density profiles have been explored for a range of applications, including plasma beam dumps~\cite{Jakobsson2019} and proton-driven wakefield acceleration in Project AWAKE~\cite{Adli2018}. While techniques such as de Laval nozzles or one-sided shock nozzles~\cite{Rovige2020} can generate steep density gradients, achieving down-ramps as short as those predicted by the optimization may remain beyond current experimental capabilities. Nevertheless, shortening the down-ramp seems to be a good strategy to minimize the time electrons spend in the decelerating phase.

The laser parameters considered in this study are optimistic, as, to the best of our knowledge, no currently available systems can provide pulse energies of several hundreds of \si{\milli\joule} at kHz repetition rates, as assumed here. For instance, EKSPLA~\cite{ekspla2024} offers an 8~\si{\femto\second}, 5~\si{\tera\watt} laser operating at 1~\si{\kilo\hertz}, delivering 42.6~\si{\milli\joule} per pulse. Similarly, Amplitude~\cite{amplitude2024} provides a 20~\si{\femto\second}, 1~\si{\tera\watt} laser capable of 20~\si{\milli\joule} per pulse at the same repetition rate. With continued progress in laser technology, it is anticipated that systems with capabilities approaching those assumed in this work will become available in the near future. An example of this progress is the recent demonstration of continuous electron-beam acceleration using a Yb:YAG laser, which delivered pulses with energies of tens of \si{\milli\joule} at repetition rates of a few \si{\kilo\hertz} \cite{Farace2026}.

\begin{acknowledgments}
The authors acknowledge the National Laboratory for Scientific Computing (LNCC/MCTI, Brazil) for providing access to the Santos Dumont supercomputer (Project LPA-FARMA), which contributed to the research results reported in this paper. This work was supported by the Conselho Nacional de Desenvolvimento Científico e Tecnológico (CNPq) under Grant 405143/2021-4 and Scholarship 140941/2023-1, and by the Fundação de Amparo à Pesquisa do Estado do Rio Grande do Sul (FAPERGS) under Grant 24/2551-0001552-3.

\end{acknowledgments}

\section*{AUTHOR DECLARATIONS\\Conflict of Interest}
The authors have no conflicts to disclose.

\section*{Author Contributions}
\textbf{B. S. Nunes:} Conceptualization (equal); Data curation (lead); Formal analysis (lead); Investigation (equal); Methodology (equal); Resources (equal); Software (lead); Validation (equal); Visualization (equal); Writing – original draft (lead); Writing – review \& editing (equal). \textbf{N. D. Vieira, Jr.:} Conceptualization (equal); Funding acquisition (equal); Writing – review \& editing (equal). \textbf{M. S. Alva-Sánchez:} Funding acquisition (equal); Writing – review \& editing (equal). \textbf{A. Bonatto:}Conceptualization (equal); Data curation (equal); Formal analysis (equal); Funding acquisition (equal); Investigation (equal); Methodology (equal);  Resources (lead); Software (equal); Supervision (equal); Validation (equal); Visualization (equal); Writing – review \& editing (equal). \textbf{R. E. Samad:} Conceptualization (equal); Data curation (equal); Formal analysis (equal); Funding acquisition (lead); Investigation (equal); Methodology (equal);  Resources (lead); Software (equal); Supervision (equal); Validation (equal); Visualization (equal); Writing – review \& editing (equal). Project administration (lead).
\section*{Data Availability Statement}
The data that support the findings of this study are available from the corresponding authors upon reasonable request.

\bibliography{referencias}

\end{document}